\documentclass[10pt,letterpaper]{article}
\usepackage[margin=1in]{geometry}
\usepackage{dcolumn}% Align table columns on decimal point
\usepackage{bm}% bold math
\usepackage{hyperref}% add hypertext capabilities
\usepackage{graphicx} % Required for inserting images
\usepackage{amsmath}
\usepackage{amssymb}
\usepackage{subcaption}
\usepackage{color}
\usepackage{graphics}
\usepackage[numbers,sort&compress]{natbib}
\usepackage{authblk}
\usepackage{titlesec}
\usepackage{booktabs}

\renewcommand{\thesection}{\Roman{section}}
\renewcommand{\thesubsection}{\Alph{subsection}}
\titleformat{\section}{\centering\normalfont\small\bfseries\MakeUppercase}
  {\thesection.}{0.5em}{}
\titleformat{\subsection}{\normalfont\small\bfseries}
  {\thesubsection.}{0.5em}{}
\titlespacing*{\section}{0pt}{2.2ex plus 0.8ex minus 0.4ex}{1.2ex}
\titlespacing*{\subsection}{0pt}{1.8ex plus 0.6ex minus 0.3ex}{0.8ex}
\hypersetup{colorlinks=true,citecolor=blue,linkcolor=blue,urlcolor=blue}

\begin{document}

\title{Quasinormal Modes Criticality across Black Hole Extremality}
\author[1]{Zi-Yu Hou}
\author[1,4]{Yu-Qi Lei\thanks{corresponding author: yuqi\_lei@shu.edu.cn}}
\author[1]{Xian-Hui Ge\thanks{corresponding author: gexh@shu.edu.cn}}
\author[1,2,3]{Bum-Hoon Lee\thanks{corresponding author: bhl@sogang.ac.kr}}
\author[1]{Li-Jun Tian\thanks{corresponding author: tianlijun@shu.edu.cn}}
\affil[1]{Department of Physics, College of Sciences, Shanghai University, 99 Shangda Road, 200444 Shanghai, China}
\affil[2]{Center for Quantum Spacetime, Sogang University, Seoul 04107, Korea}
\affil[3]{Department of Physics, Sogang University, Seoul 04107, Korea}
\affil[4]{Department of Mathematics, College of Sciences, Shanghai University, 99 Shangda Road, 200444 Shanghai, China}
\date{}
\renewcommand{\Authfont}{\normalsize}
\renewcommand{\Affilfont}{\small}
\setlength{\affilsep}{0.35em}

\maketitle
\begingroup
\endgroup

\begin{abstract}
We investigate the extremal solution branch to nonextremal branch transition in Einstein-Euler-Heisenberg anti-de Sitter black holes, where the nonlinear electromagnetic parameter $\lambda$ governs the existence of the extremal branch solution. At fixed pressure and charge, extremal black hole only exist for $\lambda\leq\lambda_c$. For $\lambda>\lambda_c$, the lowest Hawking temperature keeps nonzero. Near this threshold, the lowest temperature follows a linear critical scaling, $T_{\min}\propto(\lambda-\lambda_c)^{\alpha_T}$. We investigate the corresponding dynamical behavior by computing the quasinormal modes of a massless scalar perturbation along the lowest-temperature black hole branch. As the critical threshold is approached, two complex-frequency quasinormal modes branches collide on the imaginary axis and reorganize into distinct purely imaginary branches. Then the imaginary parts of these modes, together with higher overtones, reach zero at the critical threshold, described by $\Gamma_n\propto(\lambda-\lambda_c)^{\alpha_n}$, with $\alpha_n\simeq\alpha_T$. The resulting zero-damped modes become increasingly localized near the event horizon, and their relative damping rates closely follow the overtone spacing predicted by the near-horizon AdS$_2$ spectrum. Our results show that the thermodynamic and dynamical critical behavior remain closely connected even when extremal black hole solutions cease to exist, providing further evidence for the relevance of near-extremal spectral behavior in this distinct critical regime.
\end{abstract}
%\vspace{1em}

\section{Introduction}

The fundamental connection between black hole mechanics and thermodynamics is firmly established by the entropy-area relation and the generalized second law~\cite{Bekenstein1973BlackHolesEntropy,Bekenstein1974GeneralizedSecondLaw,Bardeen1973FourLaws}. Hawking subsequently demonstrated that black holes emit thermal radiation, establishing the physical basis of black hole temperature and its relation to surface gravity~\cite{Hawking1974BlackHoleExplosions,Hawking1975ParticleCreation}. These developments laid the foundation for subsequent studies of black hole thermodynamic phase structures and phase transitions~\cite{Hawking1983ThermodynamicsBlackHoles,Kubiznak2012BlackHoleChemistry}. Extremal and near-extremal black holes are also of particular interest, as their thermodynamic properties can exhibit distinct limiting behavior depending on the underlying gravitational theory and black hole solution~\cite{Wald1997Nernst,Wald2001Thermodynamics,GibbonsMaeda1988DilatonBlackHoles,GarfinkleHorowitzStrominger1991ChargedBlackHoles,YuGe2022ChargedDilatonPageCurves}. The diversity of thermodynamic behavior in the extremal limit also highlights the subtleties of the third law of black hole mechanics, particularly the attainability of extremality through physical processes~\cite{Israel1986ThirdLaw}. Near-extremal black holes provide a useful setting for investigating low-temperature thermodynamics and the behavior of black holes approaching extremality~\cite{HorowitzStrominger1996NearExtremal,Johnstone2013Extremal}. These studies generally consider black hole families that admit an extremal limit. A different situation arises when varying a parameter causes the extremal solutions to disappear. This raises the question of how the low-temperature thermodynamic structure changes across such a threshold and whether the corresponding nonextremal black holes retain the characteristic dynamical features of conventional near-extremal black holes. Understanding this behavior may provide further insight into the relationship between black hole thermodynamics and dynamics near extremality.

Quasinormal modes (QNMs) describe the characteristic response of black holes to linear perturbations~\cite{Kokkotas:1999bd,Berti:2009kk}. Their spectra encode information about the underlying black hole spacetime and provide an important tool for studying black hole dynamics, see Ref.~\cite{Konoplya:2011qq} for review. QNMs have also been used to investigate the quantization of black hole horizon area and entropy~\cite{Kunstatter:2002pj,Maggiore:2007nq}, further motivating studies of the interplay between black hole dynamics and thermodynamics.
An early study suggested a possible correspondence between QNMs and the second-order thermodynamic phase transition of Reissner-Nordstr\"om (RN) black holes~\cite{JingPan2008QNMsPhaseTransition}, but later work questioned whether this relation is universal for Kerr and Schwarzschild anti-de Sitter (AdS) black holes and for different perturbation sectors of RN black holes~\cite{BertiCardoso2008QNMsThermodynamicPhaseTransitions}. The QNM frequencies were found to vary differently along different thermodynamic branches, and exhibit different trends across the phase transition, in RN-AdS black holes~\cite{Liu2014}, charged AdS black holes with Weyl corrections~\cite{Mahapatra2016}, charged AdS dilaton black holes~\cite{PhysRevD.97.026014}, and regular black holes, including Bardeen and Hayward AdS black holes~\cite{LAN2021115539,GUO2024138801}. Related QNM behavior has also been found in the supercritical crossover region of RN-AdS black holes~\cite{2025arXiv250404995Z}. In \(F(R)\)-Euler-Heisenberg gravity, changes in the QNM slope were found to correspond closely to changes in the thermodynamic phase structure~\cite{HouLeiGe2026QNMsNLEDPhaseTransitions}.
These studies provide evidence that QNM spectra can reflect black hole thermodynamic behavior, although the correspondence depends on the underlying black hole background and perturbation sector. 

These developments motivate further investigation of the connection between QNMs and thermodynamics in the low-temperature regime of extremal and near-extremal black holes. Early studies of rapidly rotating Kerr black holes revealed long-lived resonances near extremality~\cite{Detweiler1980ResonantFrequencies}. These modes were subsequently understood in terms of zero-damped modes (ZDMs), whose damping rates vanish in the extremal limit and which form a distinct branch of the QNM spectrum~\cite{Yang2013BranchingQNMs,Yang2013NearExtremalKerr}. Similar ZDMs were later identified in charged near-extremal black holes, including RN and Kerr-Newman backgrounds~\cite{ZimmermanMark2016ChargedZDM}. For RN-AdS black holes, approaching extremality can also lead to a change in the dominant QNM from oscillatory to purely damped behavior~\cite{FicekWarnick2024RNAdSQNMs}. These results indicate that the QNM spectrum undergoes characteristic changes as black holes approach extremality, providing a basis for examining its relation to low-temperature thermodynamics. A different situation arises when extremal solutions disappear as a parameter crosses a critical threshold. It remains unclear whether the nonextremal black holes near this threshold retain the characteristic features of near-extremal black hole QNMs and how these features are related to the thermodynamic critical behavior.

To explore the thermodynamic and dynamical properties near the threshold where extremal solutions disappear, we consider the Einstein-Euler-Heisenberg (EEH)-AdS black hole, in which nonlinear vacuum-polarization effects modify the horizon structure and thermodynamic properties~\cite{HeisenbergEuler1936Consequences,DittrichGies2000QuantumVacuum,YajimaTamaki2001EulerHeisenbergBH,RuffiniWuXue2013EEH,MagosBreton2020EEHAdS,Myung2025Thermodynamic,HouLeiGe2026QNMsNLEDPhaseTransitions,GuoMyung2026MultipleHorizons}. A distinctive feature of the EEH-AdS black hole is that varying the EH parameter $\lambda$ can eliminate extremal black hole solutions beyond a critical threshold, rising a finite lowest Hawking temperature. This differs from the conventional near-extremal setting, in which a black hole approaches an existing extremal solution at fixed model parameters. We investigate the thermodynamic critical behavior near the threshold separating regions with and without extremal solutions, together with the QNM spectrum of a massless scalar perturbation along the corresponding lowest-temperature black hole branch. Particular attention is given to the spectral reorganization, the emergence of zero-damped modes, and the critical behavior of them. By comparing these dynamical features with the thermodynamic results, we examine whether the characteristic near-extremal behavior persists when no exact extremal solution exists at the same parameter values, and how this behavior is related to the thermodynamic criticality of the black hole.

The remainder of this paper is organized as follows. In Section~\ref{sec2}, we introduce the geometry of the EEH-AdS black hole and discuss its thermodynamic properties in the extended phase space. In Section~\ref{sec3}, we analyze the thermodynamics across the extremal-to-nonextremal transition, and derive the scaling of the lowest Hawking temperature near the critical point. In Section~\ref{sec4}, we study the quasinormal-mode spectrum of a neutral massless scalar perturbation along the corresponding low-temperature black hole branch. We demonstrate the critical scaling of the near-horizon QNM damping rates, identify the emergence of the zero-damped modes, and show that its scaling exponent agrees with that of the lowest Hawking temperature. We also analyze the near-horizon $\mathrm{AdS}_{2}$ interpretation of the QNM spectrum and the radial localization of the corresponding eigenfunctions. We summarize our conclusions in Section~\ref{sec5}. Appendix~\ref{app.A} provides a complementary analysis of the thermodynamic phase structures at fixed pressure and charge, together with the associated photon-sphere Lyapunov exponent.

\section{Extended Phase-Space Thermodynamics of the EEH-AdS Black Hole}\label{sec2}

We briefly review the four-dimensional Einstein-Euler-Heisenberg (EEH) AdS black hole and establish the extended thermodynamic framework used below. Euler-Heisenberg electrodynamics describes the leading nonlinear quantum-electrodynamic corrections to classical Maxwell theory arising from vacuum polarization effects~\cite{HeisenbergEuler1936Consequences}. Throughout this work, we adopt natural units, $G=\hbar=c=k_{\mathrm B}=1$. The corresponding gravitational action coupled to Euler-Heisenberg nonlinear electrodynamics can be written as~\cite{MagosBreton2020EEHAdS}
\begin{equation}\label{eq:eeh-action}
\begin{aligned}
&I=\frac{1}{4\pi}\int \mathrm{d}^{4}x\sqrt{-g}\left[\frac{R-2\Lambda}{4}-\mathcal{L}(X,Y)\right],\\
&\mathcal{L}(X,Y)=-X+\frac{\lambda}{2}X^{2}+\frac{7\lambda}{8}Y^{2},
\end{aligned}
\end{equation}
where $R$ is the Ricci scalar, $\Lambda=-3/l^{2}$ is the negative cosmological constant with $l$ denoting the AdS curvature radius, and $\lambda$ is the Euler-Heisenberg parameter. The electromagnetic invariants are
\begin{equation}\label{eq:em-invariants}
X=\frac{1}{4}F_{\mu\nu}F^{\mu\nu}=\frac{\mathbf{B}^{2}-\mathbf{E}^{2}}{2},
\qquad
Y=\frac{1}{4}{}^{*}F_{\mu\nu}F^{\mu\nu}=\mathbf{E}\cdot\mathbf{B}.
\end{equation}

Static and spherically symmetric black hole solutions in Einstein-Euler-Heisenberg theory have been studied in Refs.~\cite{YajimaTamaki2001EulerHeisenbergBH,RuffiniWuXue2013EEH}. For a purely electric configuration, for which $Y=0$, the corresponding AdS solution takes the form~\cite{MagosBreton2020EEHAdS}
\begin{equation}\label{eq:eeh-metric}
\begin{aligned}
\mathrm{d}s^{2}&=-f(r)\mathrm{d}t^{2}+\frac{\mathrm{d}r^{2}}{f(r)}+r^{2}\mathrm{d}\Omega_{2}^{2},\\
f(r)&=1-\frac{2M}{r}+\frac{r^{2}}{l^{2}}+\frac{Q^{2}}{r^{2}}-\frac{\lambda Q^{4}}{20r^{6}},
\end{aligned}
\end{equation}
where $M$ and $Q$ denote the black hole mass and electric charge, respectively. The event horizon $r_{+}$ is the largest positive root of $f(r)=0$, from which the mass can be expressed as
\begin{equation}\label{eq:mass-l}
M=\frac{r_{+}}{2}+\frac{Q^{2}}{2r_{+}}-\frac{\lambda Q^{4}}{40r_{+}^{5}}+\frac{r_{+}^{3}}{2l^{2}}.
\end{equation}
The Hawking temperature is determined by the surface gravity at the event horizon, while the entropy obeys the Bekenstein-Hawking area law and the electric potential is defined by the potential difference between the horizon and infinity~\cite{MagosBreton2020EEHAdS},
\begin{equation}\label{eq:T}
T=\frac{f'(r_{+})}{4\pi}=\frac{3r_{+}}{4\pi l^{2}}+\frac{1}{4\pi r_{+}}-\frac{Q^{2}}{4\pi r_{+}^{3}}+\frac{\lambda Q^{4}}{16\pi r_{+}^{7}},
\end{equation}
and
\begin{equation}\label{eq:entropy-potential}
S=\pi r_{+}^{2},
\qquad
\Phi=\frac{Q}{r_{+}}\left(1-\frac{\lambda Q^{2}}{10r_{+}^{4}}\right).
\end{equation}

In extended black hole thermodynamics, the cosmological constant is promoted to a thermodynamic pressure, with its conjugate variable interpreted as the thermodynamic volume~\cite{KastorRayTraschen2009,Dolan2011PressureVolume,Kubiznak2012BlackHoleChemistry},
\begin{equation}\label{eq:pressure}
P=-\frac{\Lambda}{8\pi}=\frac{3}{8\pi l^{2}}.
\end{equation}
Within this framework, the mass of an asymptotically AdS black hole is interpreted as the thermodynamic enthalpy rather than the internal energy~\cite{KastorRayTraschen2009,Dolan2011PressureVolume}. Using the horizon condition together with Eq.~\eqref{eq:pressure}, the mass becomes
\begin{equation}\label{eq:enthalpy}
M=\frac{r_{+}}{2}+\frac{Q^{2}}{2r_{+}}-\frac{\lambda Q^{4}}{40r_{+}^{5}}+\frac{4\pi P r_{+}^{3}}{3},
\end{equation}
and may therefore be regarded as a thermodynamic function $M=M(S,Q,P,\lambda)$.

The quantities conjugate to $P$ and $\lambda$ follow directly from the enthalpy,
\begin{equation}\label{eq:extended-conjugates}
V=\left(\frac{\partial M}{\partial P}\right)_{S,Q,\lambda}=\frac{4\pi r_{+}^{3}}{3},
\qquad
\mathcal{A}=\left(\frac{\partial M}{\partial\lambda}\right)_{S,Q,P}=-\frac{Q^{4}}{40r_{+}^{5}}.
\end{equation}
For the EEH-AdS solution, these expressions agree with the corresponding thermodynamic quantities obtained from the explicit mass function~\cite{MagosBreton2020EEHAdS}.

More generally, when dimensionful coupling parameters entering a nonlinear electromagnetic Lagrangian are allowed to vary, their variations generate additional work terms in the first law of black hole mechanics~\cite{ZhangGao2018,BokulicJuricSmolic2021}. Combining this generalized nonlinear-electrodynamics framework with extended AdS thermodynamics, the EEH-AdS black hole satisfies
\begin{equation}\label{eq:extended-first-law}
\mathrm{d}M=T\mathrm{d}S+\Phi\mathrm{d}Q+V\mathrm{d}P+\mathcal{A}\mathrm{d}\lambda.
\end{equation}
For variations restricted to a fixed EEH theory, one sets $\mathrm{d}\lambda=0$. Nevertheless, because $\lambda$ is a dimensionful coupling, its scaling weight contributes to the generalized Smarr relation when the enlarged thermodynamic parameter space is considered~\cite{ZhangGao2018,GulinSmolic2018,BokulicJuricSmolic2021}.

Using Eq.~\eqref{eq:pressure}, the Hawking temperature can be expressed directly in terms of the thermodynamic pressure. Solving this relation for $P$ gives the equation of state of the EEH-AdS black hole~\cite{MagosBreton2020EEHAdS},
\begin{equation}\label{eq:temperature-eos}
\begin{aligned}
T&=2Pr_{+}+\frac{1}{4\pi r_{+}}-\frac{Q^{2}}{4\pi r_{+}^{3}}+\frac{\lambda Q^{4}}{16\pi r_{+}^{7}},\\
P&=\frac{T}{2r_{+}}-\frac{1}{8\pi r_{+}^{2}}+\frac{Q^{2}}{8\pi r_{+}^{4}}-\frac{\lambda Q^{4}}{32\pi r_{+}^{8}}.
\end{aligned}
\end{equation}

To derive the corresponding Smarr relation, we note that in four spacetime dimensions the thermodynamic variables scale as
\begin{equation}
S\to\xi^{2}S,\qquad
Q\to\xi Q,\qquad
P\to\xi^{-2}P,\qquad
\lambda\to\xi^{2}\lambda,
\end{equation}
while $M\to\xi M$. The enthalpy is therefore a weighted homogeneous function satisfying
\begin{equation}\label{eq:enthalpy-scaling}
M(\xi^{2}S,\xi Q,\xi^{-2}P,\xi^{2}\lambda)=\xi M(S,Q,P,\lambda).
\end{equation}
Applying Euler's theorem for homogeneous functions and using the thermodynamic conjugate relations gives
\begin{equation}\label{eq:smarr}
M=2TS+\Phi Q-2PV+2\lambda\mathcal{A}.
\end{equation}
The additional contribution associated with the dimensionful nonlinear-electrodynamics coupling is consistent with the generalized Smarr relations established for black holes coupled to nonlinear electromagnetic fields~\cite{GulinSmolic2018,ZhangGao2018,BokulicJuricSmolic2021}. For the specific EEH-AdS solution, Eq.~\eqref{eq:smarr} agrees with the explicit extended thermodynamic relation obtained in Ref.~\cite{MagosBreton2020EEHAdS}. Accordingly, the relations presented above constitute the specialization and explicit verification of the general extended black hole thermodynamic framework for the EEH-AdS solution.

\section{Thermodynamics across the Extremal-to-Nonextremal Transition}\label{sec3}

The nonlinear term in Euler-Heisenberg theory affects the existence of extremal black holes and their thermodynamic structure. In this section, we investigate the associated extremal-to-nonextremal transition, with attention to its critical behavior. To characterize this transition, we fix the thermodynamic pressure $P$ and charge $Q$ and vary the Euler-Heisenberg parameter $\lambda$. The existence of the extremal black hole solution is then determined by the zero-temperature structure of the Hawking temperature $T(r_{+},P,Q,\lambda)$.

At fixed $P$ and $Q$, the critical threshold is reached when the local minimum of the temperature function approaches zero from the nonextremal side. The corresponding critical conditions are
\begin{equation}
T(r_c,P,Q,\lambda_c)=0,\qquad
\left.\frac{\partial T}{\partial r_+}\right|_c=0,
\label{eq:eeh-critical-conditions}
\end{equation}
where the subscript $c$ denotes evaluation at the critical point. Substituting Eq.~\eqref{eq:temperature-eos} into Eq.~\eqref{eq:eeh-critical-conditions} and eliminating $\lambda_{c}$ yields
\begin{equation}
32\pi P r_{c}^{4}+3r_{c}^{2}-2Q^{2}=0.
\label{eq:eeh-critical-radius-polynomial}
\end{equation}
The physically admissible positive root and the corresponding critical Euler-Heisenberg parameter are
\begin{equation}
\begin{aligned}
r_{c}(P,Q)
&=\left(
\frac{-3+\sqrt{9+256\pi P Q^{2}}}{64\pi P}
\right)^{1/2},
\\
\lambda_{c}(P,Q)
&=\frac{r_{c}^{4}\left(2Q^{2}-r_{c}^{2}\right)}{Q^{4}}.
\end{aligned}\label{eq:eeh-critical}
\end{equation}

\begin{figure}[htbp]
\centering
\includegraphics[width=\linewidth]{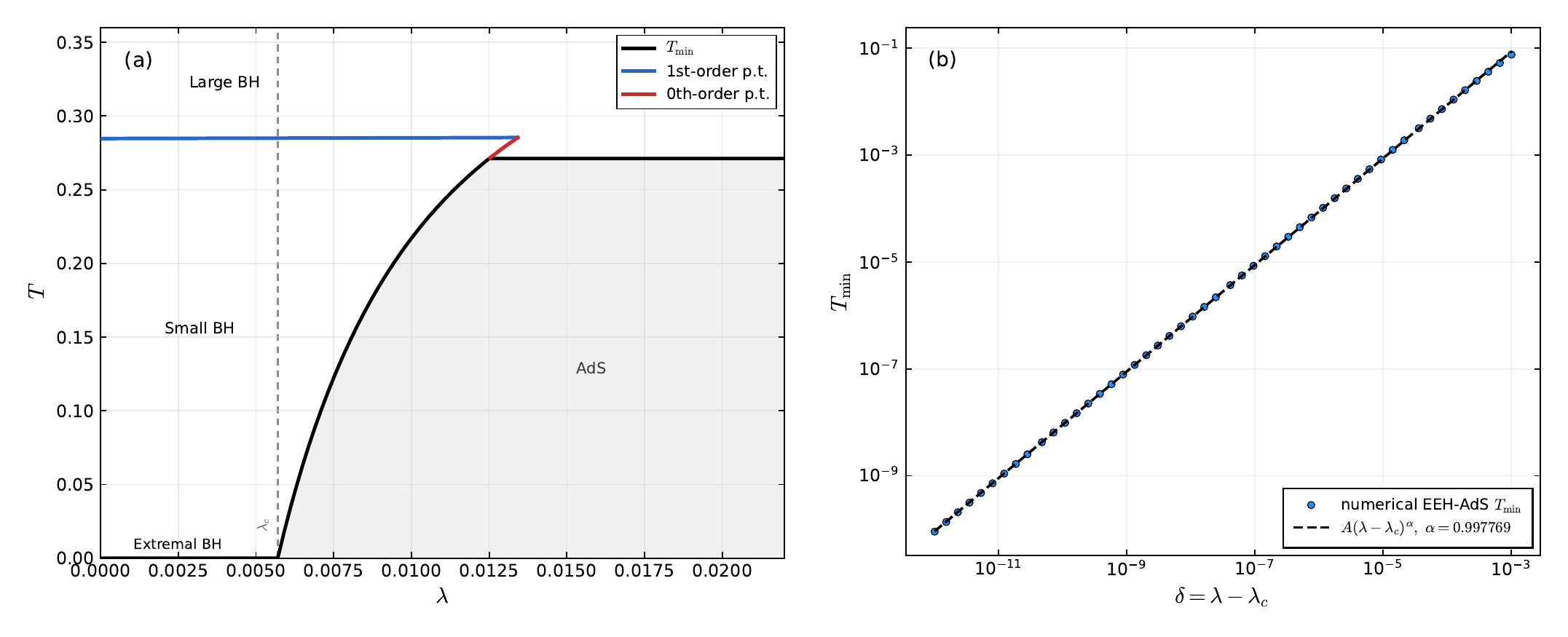}
\caption{Thermodynamic phase structure of the EEH-AdS black hole and the critical behavior of the extremal-to-nonextremal transition. (a) The $T$-$\lambda$ phase diagram at fixed charge and pressure. (b) Log-log fit of the lowest Hawking temperature $T_{\min}$ as a function of $\delta=\lambda-\lambda_c$ near the critical coupling. The numerical results satisfy $T_{\min}=A(\lambda-\lambda_c)^\alpha$, with $\alpha=0.997769\simeq1$.}
\label{fig:eeh-lowest-temperature-fit}
\end{figure}

To examine the black hole configurations on both sides of the critical point, we fix $P$ and $Q$ and vary the EH parameter $\lambda$. For $\lambda<\lambda_c$, the EEH-AdS black hole admits an extremal configuration with a degenerate outer event horizon. At $\lambda=\lambda_c$, the limiting extremal configuration is characterized by the critical horizon radius $r_c$. For $\lambda>\lambda_c$, no extremal black hole solution exists, and the lowest Hawking temperature becomes strictly positive. The physical black hole branch then satisfies
\begin{equation}
T\geq T_{\min}>0.
\end{equation}
Therefore, $\lambda_c(P,Q)$ marks the threshold separating parameter regions with and without extremal black hole solutions. This behavior is illustrated in Fig.~\ref{fig:eeh-lowest-temperature-fit}(a). A more detailed discussion of the thermodynamic phase structure is provided in Appendix~\ref{app.A}, and some other related aspects can be found in Refs.~\cite{MagosBreton2020EEHAdS,Ye:2022uuj,Zhao:2024phz}.  

To characterize the approach to the critical set from the nonextremal side, we introduce
\begin{equation}
\delta\equiv\lambda-\lambda_{c}(P,Q)>0,
\qquad
r_{\min}=r_{c}(P,Q)+\epsilon,
\label{eq:eeh-critical-distance}
\end{equation}
where $r_{\min}$ denotes the location of the relevant local minimum of the Hawking temperature. Expanding the condition for a local extremum, $\partial T/\partial r_{+}=0$, around the critical point gives
\begin{equation}
0=
\left.\frac{\partial^{2}T}{\partial r_{+}^{2}}\right|_{c}\epsilon
+\left.\frac{\partial^{2}T}{\partial r_{+}\partial\lambda}\right|_{c}\delta
+\mathcal{O}\left(\epsilon^{2},\epsilon\delta,\delta^{2}\right).
\label{eq:eeh-stationary-expansion}
\end{equation}
Provided that the critical point is nondegenerate,
$\left.\partial^{2}T/\partial r_{+}^{2}\right|_{c}\neq0$, the displacement of the minimum is
\begin{equation}
\epsilon
=-\left.
\frac{\partial^{2}T/\partial r_{+}\partial\lambda}
{\partial^{2}T/\partial r_{+}^{2}}
\right|_{c}\delta
+\mathcal{O}(\delta^{2}).
\label{eq:eeh-minimum-radius-shift}
\end{equation}
Thus, $\epsilon=\mathcal{O}(\delta)$ near the critical point.

This result also determines how the lowest temperature emerges on the nonextremal side. Since the critical point satisfies
$\left.T\right|_{c}=0$ and
$\left.\partial T/\partial r_{+}\right|_{c}=0$, the displacement $\epsilon$ does not contribute to $T_{\min}$ at linear order. Because $\epsilon=\mathcal{O}(\delta)$, all contributions associated with the shift of the minimum begin at order $\delta^{2}$. The leading variation of $T_{\min}$ is therefore controlled by the explicit dependence of $T$ on $\lambda$. Expanding the temperature at the shifted minimum gives
\begin{align}
T_{\min}
&=T(r_{c}+\epsilon,P,Q,\lambda_{c}+\delta)
\nonumber\\
&=\left.\frac{\partial T}{\partial\lambda}\right|_{c}\delta
+\mathcal{O}(\delta^{2}).
\label{eq:eeh-lowest-temperature-expansion}
\end{align}
For the EEH-AdS black hole, this coefficient is nonzero,
\begin{equation}
\left.\frac{\partial T}{\partial\lambda}\right|_{c}
=\frac{Q^{4}}{16\pi r_{c}^{7}}\neq0.
\label{eq:eeh-temperature-lambda-derivative}
\end{equation}
Consequently, the lowest Hawking temperature grows linearly with the distance from the critical point on the nonextremal side,
\begin{equation}
T_{\min}\sim(\lambda-\lambda_{c})^{\alpha_{T}},
\qquad
\alpha_{T}=1.
\label{eq:eeh-thermal-critical-exponent}
\end{equation}
The unit critical exponent therefore follows directly from the nonvanishing linear response of the temperature to $\lambda$ at the critical point.

As shown in Fig.~\ref{fig:eeh-lowest-temperature-fit}(b), the numerical results confirm this linear behavior. The fitted exponent $\alpha_{\mathrm{fit}}=0.997769$ agrees closely with the analytical prediction $\alpha_T=1$. The deviation from unity is consistent with subleading corrections over the finite fitting interval.

The critical EH parameter $\lambda_c(P,Q)$ defines the threshold beyond which extremal black hole solutions cease to exist. At this threshold, the limiting extremal horizon is located at $r_+=r_c(P,Q)$. Changing $P$ or $Q$ shifts the critical values $r_c$ and $\lambda_c$ and modifies the coefficient of the linear relation between $T_{\min}$ and $\lambda-\lambda_c$. However, as long as the critical point remains nondegenerate and $\left.\partial T/\partial\lambda\right|_c\neq0$, the critical exponent remains $\alpha_T=1$. %Thus, the emergence of a finite lowest Hawking temperature on the nonextremal side is characterized by the same linear critical behavior.

\section{Quasinormal Modes across the Extremal-to-Nonextremal Transition}\label{sec4}

Having established the critical thermodynamic scaling near the extremal-to-nonextremal threshold, we now investigate its dynamical imprint. We examine whether the macroscopic thermodynamic scale $T_{\mathrm{min}}$ governs the relaxation dynamics of the spacetime. To probe this, we evaluate the near-horizon QNMs of a massless scalar perturbation on the EEH-AdS background. 

%%%%%%%%
\subsection{Numerical results of QNMs}

The scalar field satisfies the massless Klein-Gordon equation,
\begin{equation}
\Box\Psi
=\frac{1}{\sqrt{-g}}
\partial_{\mu}\!\left(
\sqrt{-g}\,g^{\mu\nu}\partial_{\nu}\Psi
\right)=0.
\label{eq:eeh-klein-gordon}
\end{equation}
We introduce the ingoing Eddington-Finkelstein coordinate
\begin{equation}
v=t+r_{*},
\qquad
\frac{\mathrm{d}r_{*}}{\mathrm{d}r}=\frac{1}{f(r)},
\label{eq:eeh-ef-coordinate}
\end{equation}
where $r_{*}$ is the tortoise coordinate and $f(r)$ is the blackening function. With the compact coordinate
\begin{equation}
u=\frac{r_{+}}{r},
\end{equation}
the event horizon and the asymptotic AdS boundary are located at $u=1$ and $u=0$, respectively. We separate variables according to
\begin{equation}
\Psi(v,u,\theta,\varphi)
=e^{-i\omega v}Y_{\ell m}(\theta,\varphi)
u^{3}\widetilde{\phi}(u),
\label{eq:eeh-scalar-ansatz}
\end{equation}
where the spherical harmonics satisfy
\begin{equation}
\nabla_{S^{2}}^{2}Y_{\ell m}
=-\ell(\ell+1)Y_{\ell m}.
\label{eq:eeh-spherical-harmonics}
\end{equation}
Defining
\begin{equation}
\tilde{f}(u)\equiv f\!\left(\frac{r_{+}}{u}\right),
\end{equation}
substitution of Eq.~\eqref{eq:eeh-scalar-ansatz} into Eq.~\eqref{eq:eeh-klein-gordon} yields
\begin{align}
&u^{2}\tilde{f}(u)\widetilde{\phi}''(u)
+\left[
u^{2}\tilde{f}'(u)+6u \tilde{f}(u)+2i\omega r_{+}
\right]\widetilde{\phi}'(u)
\nonumber\\
&\quad
+\left[
3u \tilde{f}'(u)+6\tilde{f}(u)-\ell(\ell+1)
+\frac{4i\omega r_{+}}{u}
\right]\widetilde{\phi}(u)=0.
\label{eq:eeh-qnm-radial}
\end{align}

The ingoing behavior at the event horizon is already incorporated through the factor $e^{-i\omega v}$. The radial function $\widetilde{\phi}(u)$ is therefore required to be regular at $u=1$. At the AdS boundary, the source-free boundary condition is implemented by requiring \(\widetilde{\phi}(u)\) to remain finite as \(u\to0\). In the numerical analysis below, we employ the Chebyshev pseudo-spectral method to solve the radial equation with $N+1=321$ collocation points and set $\ell=0$.

%%%%%%%%%%

As in the thermodynamic analysis, we define
\begin{equation}
\delta\equiv\lambda-\lambda_{c}(P,Q)>0.
\label{eq:eeh-qnm-critical-distance}
\end{equation}
Throughout the numerical analysis, $P$ and $Q$ are held fixed. For each $\lambda>\lambda_c$, the QNM spectrum is evaluated on the lowest temperature black hole solution at $r_{+}=r_{\min}(\lambda)$. Thus, as $\delta\to0^{+}$, the QNM calculation follows the same family of black hole backgrounds used to determine the thermodynamic critical behavior. This makes $\delta$ a direct link between the two analyses: both $T_{\min}$ and the QNM spectrum can be tracked as the same critical point is approached.

\begin{figure}[htbp]
\centering
\includegraphics[width=\linewidth]{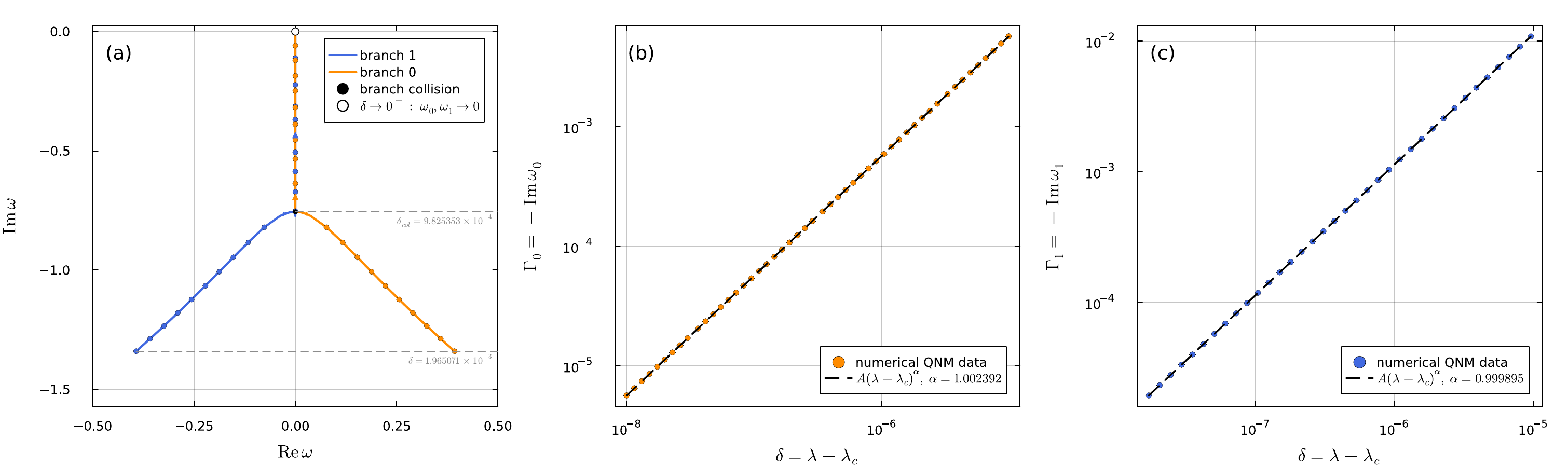}
\caption{
Critical evolution and scaling of the $\ell=0$ QNMs near the extremal-to-nonextremal threshold.
(a) Two QNM branches approach each other as $\delta=\lambda-\lambda_{c}$ decreases, meet and separate into two purely imaginary branches.
(b) Log-log scaling of $\Gamma_{0}=-\operatorname{Im}\omega_0$. The fit gives $\alpha_{0}=1.002392\simeq1$. 
(c) Log-log scaling of $\Gamma_{1}=-\operatorname{Im}\omega_1$. The fit gives $\alpha_{1}=0.999895\simeq1$. 
}
\label{fig:eeh-qnm-branches-scaling}
\end{figure}

The QNM spectrum exhibits a clear dynamical signature of the thermodynamic critical behavior identified above. As the critical limit is approached from the nonextremal side, the two complex frequency branches move toward the imaginary axis and undergo a spectral collision. Before the collision, the two modes are symmetric with respect to the imaginary axis and satisfy
$\omega_{1}=-\omega_{0}^{*}$.
At the collision, their real parts vanish, after which the spectrum reorganizes into two distinct purely imaginary branches. This transition marks a qualitative change in the QNM spectrum as the thermodynamic critical point is approached.

The subsequent evolution of the purely imaginary modes reveals the emergence of zero-damping behavior. We identify the two branches with the lowest modes, $n=0$ and $n=1$. Although their trajectories immediately after the collision are different, both modes ultimately move toward the origin as $\delta\to0^{+}$, so that
$\omega_{0},\omega_{1}\to0$ and hence
$-\operatorname{Im}\omega_{n}\to0$.
The vanishing imaginary parts imply increasingly long-lived perturbations and are characteristic of zero-damped modes (ZDMs)~\cite{Yang2013BranchingQNMs,Yang2013NearExtremalKerr,ZimmermanMark2016ChargedZDM}. The spectral collision and the subsequent formation of purely imaginary ZDM branches therefore constitute a dynamical spectral reorganization associated with the approach to the extremal-to-nonextremal critical point.

To quantify the near-critical damping behavior, we define
\begin{equation}
\Gamma_n\equiv-\operatorname{Im}\omega_n>0,
\label{eq:eeh-qnm-damping-rate}
\end{equation}
and fit the damping rates to
\begin{equation}
\Gamma_n=A_n\delta^{\alpha_n}.
\label{eq:eeh-qnm-branch-scaling}
\end{equation}
Figures~\ref{fig:eeh-qnm-branches-scaling}(b) and \ref{fig:eeh-qnm-branches-scaling}(c) show the corresponding logarithmic fits for the two purely imaginary branches, both yielding exponents close to unity. The analysis is extended to the first five resolved modes, with the fitted exponents and relative damping rates summarized in Table~\ref{tab:eeh-qnm-critical-exponents}.

%%%%%%%%%%%

The linear scaling is not restricted to the two modes. Table~\ref{tab:eeh-qnm-critical-exponents} lists independent fits for the near-horizon modes with overtone numbers $n=0,\ldots,4$. For each mode, the damping rate is fitted to $\Gamma_{n}=A_{n}\delta^{\alpha_{n}}.$ All fitted exponents remain close to unity, with deviations smaller than approximately $0.24\%$.

\begin{table}[htbp]
\centering
\caption{Critical scaling of the first five resolved $\ell=0$ near-horizon QNMs. The final column gives the median damping-rate ratio evaluated at common values of $\delta$.}
\label{tab:eeh-qnm-critical-exponents}
\begin{tabular}{ccccc}
\toprule
$n$ & $\alpha_{n}$ & $R^{2}$ & $\Gamma_{n}/\Gamma_{0}$ \\
\midrule
$0$ & $1.002392$ & $0.9999990$ & $1$ \\
$1$ & $0.999895$ & $0.9999999$ & $1.99087$ \\
$2$ & $1.002196$ & $0.9999995$ & $2.98874$ \\
$3$ & $0.999226$ & $0.9999986$ & $3.95857$ \\
$4$ & $1.002399$ & $0.9999978$ & $4.96353$ \\
\bottomrule
\end{tabular}
\end{table}

The first five resolved modes therefore exhibit the same approximately linear critical behavior. Since the thermodynamic analysis gives $\alpha_{T}=1$, the numerical results further indicate
\begin{equation}
\alpha_{n}^{\mathrm{QNM}}\simeq\alpha_{T}=1,
\qquad n=0,\ldots,4,
\label{eq:eeh-qnm-thermal-exponent-matching}
\end{equation}
within numerical accuracy. The simultaneous vanishing of $T_{\min}$ and of the damping rates of the first five resolved near-horizon modes therefore provides consistent thermodynamic and dynamical signatures of the same critical threshold. The relative spacing of these modes and its implication for the ZDM structure will be examined below.
 
%%%%%%%%%%%
\subsection{Comparison with the Near-Horizon AdS$_2$ Description}

The near critical QNM spectrum can be understood from the near-horizon geometry of the nearly extremal black hole. As the inner and outer horizons approach each other, the blackening function in the throat region takes the form
\begin{equation}
f(r)\simeq
\frac{(r-r_{+})(r-r_{-})}{L_{2}^{2}},
\label{eq:hhhrizon}
\end{equation}
where $L_{2}$ is the characteristic AdS$_2$ scale. The corresponding near-horizon region approaches a finite-temperature $\mathrm{AdS}_{2}\times S^{2}$ geometry, whose temperature tends to zero in the extremal limit.

In the near-extremal and low-frequency regime, the near-horizon poles of a neutral scalar perturbation take the leading-order form~\cite{FaulknerLiuMcGreevyVegh2011,Ohya2013FiniteTemperatureAdS2,Zhang2021PairProductionRNAdS}
\begin{equation}
\omega_n\simeq-2\pi iT(n+\Delta),
\qquad n=0,1,2,\ldots,
\label{eq:omegan}
\end{equation}
where $\Delta$ is the effective conformal dimension. The corresponding damping rates therefore satisfy
\begin{equation}
\Gamma_n\equiv-\operatorname{Im}\omega_n
\simeq2\pi T(n+\Delta).
\label{eq:eeh-ads2-damping}
\end{equation}
Thus, each fixed-$n$ mode becomes zero damped as $T\to0$, naturally producing a tower of ZDMs.

Along the lowest-temperature branch, $T=T_{\min}(\lambda)\propto\delta^{\alpha_T}$ with $\alpha_T=1$. For modes following the standard AdS$_2$ temperature dependence, this implies $\Gamma_n\propto\delta$ and hence $\alpha_n^{\mathrm{QNM}}=1$, consistent with the numerical exponents obtained above.

For a minimally coupled scalar field,
\begin{equation}
\Delta=\frac{1}{2}
+\sqrt{\frac{1}{4}+m_{\mathrm{eff}}^{2}L_{2}^{2}},
\qquad
m_{\mathrm{eff}}^{2}
=m_s^{2}+\frac{\ell(\ell+1)}{r_c^{2}}.
\label{eq:eeh-ads2-conformal-dimension}
\end{equation}
For the massless $\ell=0$ perturbation considered here, $\Delta=1$, so that
\begin{equation}
\Gamma_n\simeq2\pi(n+1)T,
\qquad
\frac{\Gamma_n}{\Gamma_0}\simeq n+1.
\label{eq:Deltagose1}
\end{equation}
The two purely imaginary branches in Fig.~\ref{fig:eeh-qnm-branches-scaling}(a) therefore correspond to the fundamental mode and the first overtone of this near-horizon sequence. The higher-mode damping-rate ratios listed in Table~\ref{tab:eeh-qnm-critical-exponents} also follow the predicted overtone spacing closely.

The agreement in critical exponents and relative overtone spacing indicates that the near-critical QNM spectrum of the EEH-AdS black hole exhibits features consistent with the standard AdS$_2$ near-horizon description. In particular, the characteristic low-frequency behavior persists as the EH parameter approaches the threshold beyond which extremal black hole solutions cease to exist. These results suggest that familiar near-extremal spectral relations remain relevant in this distinct critical regime, although their precise geometric origin requires further investigation.

%%%%%%%%%%%%
\subsection{Near-Horizon Localization}

The spectral analysis above indicates that the low-frequency QNMs become increasingly associated with the near-horizon AdS$_2$ region as the critical limit is approached. To examine this behavior directly at the level of the eigenfunctions, we now study how the radial profiles of the two spectral branches evolve with decreasing $\delta$. In particular, we ask whether the perturbations become progressively concentrated toward the event horizon as the black hole approaches the extremal limit.

We continue to use the compact coordinate $u=r_+/r$, for which the AdS boundary is located at $u=0$ and the event horizon at $u=1$. The master variable obtained from the numerical spectral calculation is denoted by $\widetilde{\phi}_n(u)$, and the corresponding scalar radial profile is written as
\begin{equation}
\Phi_n(u)=u^3\widetilde{\phi}_n(u).
\end{equation}

Although QNM eigenfunctions lack a standard probabilistic interpretation, the weighted integral $u_{\mathrm{eff},n}$ defined in Eq.(\ref{eq:qnm-ueff}) serves as a robust positive-definite diagnostic for radial localization.
\begin{equation}
    u_{{\rm eff},n}
    =\frac{\displaystyle\int_0^1 u\,|\Phi_n(u)|^2\,\mathrm{d}u}
    {\displaystyle\int_0^1 |\Phi_n(u)|^2\,\mathrm{d}u}.
    \label{eq:qnm-ueff}
\end{equation}
According to the definition, if $u_{{\rm eff},n}$ approaches unity, the radial profile becomes increasingly concentrated in the near-horizon region. Conversely, if $u_{{\rm eff},n}$ moves away from unity, the radial profile becomes increasingly weighted toward the exterior region.

\begin{figure}[htbp]
    \centering
    \includegraphics[width=0.5\linewidth]{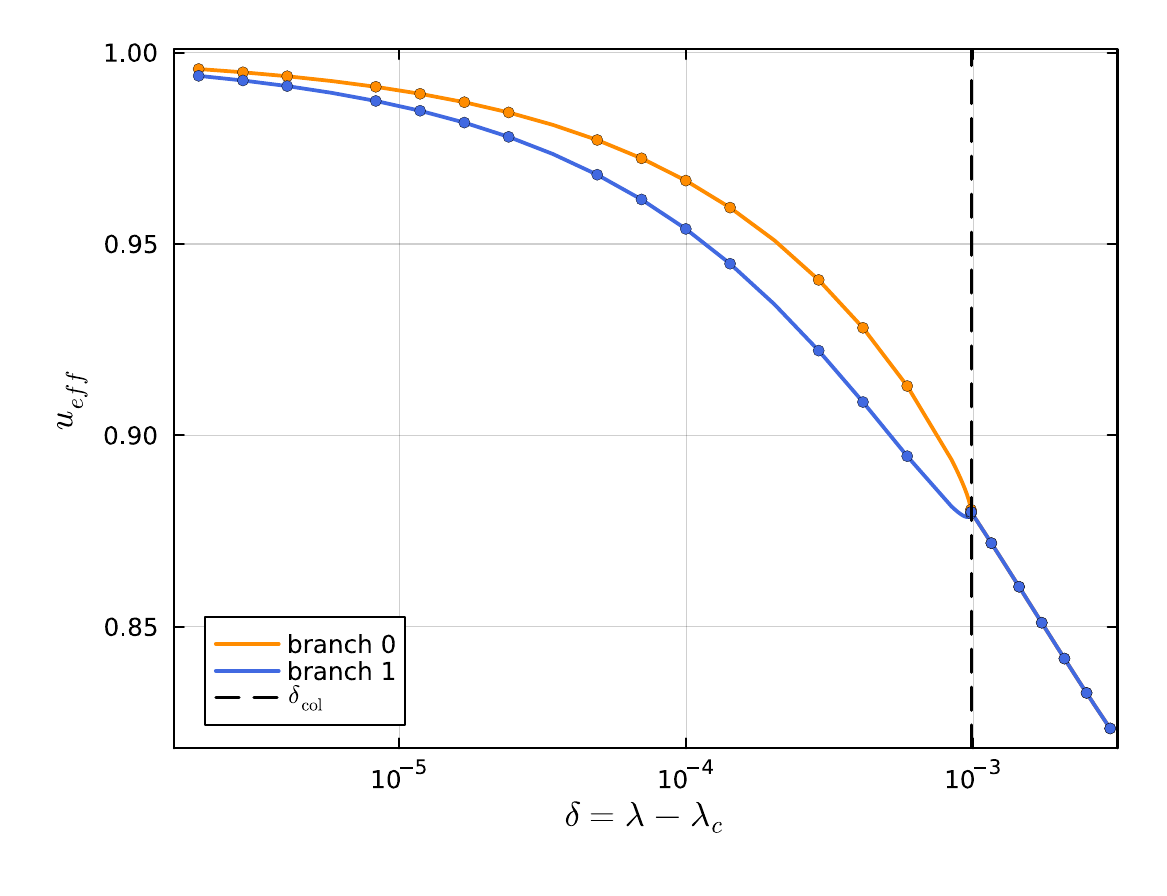}
    \caption{Effective radial position $u_{{\rm eff},n}$ calculated from the scalar radial profiles, with $Q=0.1$ and $P=3/(8\pi)$. }
    \label{fig:qnm-localization-decoupling}
\end{figure}

As shown in Fig.~\ref{fig:qnm-localization-decoupling}, the effective radial positions of the two branches coincide in the complex-frequency regime and become distinct after the spectral collision. As $\delta\to0^{+}$, however, both $u_{\mathrm{eff},n}$ approach unity, indicating that the corresponding eigenfunctions become increasingly concentrated near the event horizon. This behavior provides further evidence for the near-horizon localization of the zero-damped modes, despite the different radial structures of the two purely imaginary branches.

The evolution of $u_{{\rm eff},n}$ provides direct evidence that both purely imaginary branches become progressively localized in the near-horizon region as the extremal limit is approached. The two branches first meet on the imaginary axis and then separate into distinct purely imaginary modes with different radial structures. As $\delta$ decreases further, however, their effective radial positions gradually approach each other and both tend toward the event horizon. This eigenfunction behavior is consistent with the spectral evolution and the near-horizon interpretation inferred from the QNM analysis, further supporting the identification of these modes with the near-horizon ZDM sector.

\section{Conclusion}\label{sec5}

In summary, we investigate the thermodynamic critical behavior and its dynamical signatures near the extremal-to-nonextremal transition of the four-dimensional EEH-AdS black hole. The main result is that the lowest Hawking temperature and the damping rates of the near-critical QNMs vanish with the same linear critical behavior as the transition threshold is approached from the nonextremal side, with $\alpha_n^{\mathrm{QNM}}\simeq\alpha_T$. This agreement reveals a quantitative connection between thermodynamic criticality and QNM dynamics in a setting where extremal black hole solutions cease to exist beyond a critical value of the EH parameter.

At fixed pressure $P$ and charge $Q$, extremal black hole solutions exist for $\lambda\leq\lambda_c(P,Q)$, whereas for $\lambda>\lambda_c$ no extremal solutions exist and the lowest Hawking temperature becomes finite, increasing linearly with $\lambda-\lambda_c$. Along the corresponding lowest-temperature black hole branch, the QNM spectrum exhibits a characteristic spectral reorganization. As $\lambda$ decreases toward the critical threshold, two complex-frequency branches meet on the imaginary axis and continue as distinct purely imaginary branches. Upon approaching the threshold further, their damping rates vanish, exhibiting the zero-damped behavior familiar from nearly extremal black holes. The higher resolved modes follow the same approximately linear critical law, while their relative damping rates closely agree with the overtone spacing predicted by the AdS$_2$ near-horizon spectrum. The corresponding eigenfunctions also become increasingly localized near the event horizon. Together, these results show that the disappearance of extremal black hole solutions is accompanied by a distinct evolution of the low-frequency spectrum, characterized by mode collision, vanishing damping, and near-horizon localization.

The broader significance of these findings lies in the persistence of familiar near-extremal dynamical behavior in a setting where the existence of extremal black holes itself changes across a parameter threshold. The observed relation between the QNM damping rates and the lowest Hawking temperature, together with the near-integer overtone spacing, is consistent with the low-frequency scaling structure associated with AdS$_2$ near-horizon physics. Our results thus provide evidence that such near-horizon spectral relations can remain relevant even when the critical limit is approached through a family of nonextremal black holes without extremal counterparts at the same parameter values. This extends the evidence for the applicability of familiar near-horizon scaling behavior to a distinct type of critical process, rather than establishing a new universal law. Further investigation of the critical near-horizon geometry and other perturbation sectors will help clarify the conditions under which this correspondence persists. The observed spectral collision also motivates a more detailed examination of the eigenvalues and eigenfunctions to determine whether it is associated with an exceptional point.

\section*{Acknowledgement}
The work is supported in part by NSFC, China (Grant No. 12275166, No.12405072 and No.12675074) and China Postdoctoral Science Foundation (Grant No. 2024M761914). BHL was supported by the National Research Foundation of Korea (NRF) (RS-2020-NR049598, RS-2024-00339204, and RS-2026-25473640) and by Overseas Visiting Fellow Program of Shanghai University. BHL thanks Asia Pacific Center for Theoretical Physics (APCTP) and Korea Institute for Advanced Study (KIAS) for the hospitality during his visit, where a part of this project was done.

\appendix
\renewcommand{\thesection}{\Alph{section}}
\titleformat{\section}
  {\centering\normalfont\small\bfseries}
  {Appendix~\thesection:}{0.5em}{}
\numberwithin{equation}{section}
\numberwithin{figure}{section}
\numberwithin{table}{section}

\section{Thermodynamic Phase Structures}
\label{app.A}

In this appendix, we present the thermodynamic phase structure of the EEH-AdS black hole at fixed pressure $P$ and charge $Q$. For each value of the EH parameter $\lambda$, the black hole branches are parametrized by the horizon radius $r_+$. 

As established in Sec.~\ref{sec2}, the black hole mass and Hawking temperature are
\begin{equation}
\begin{aligned}
M&=\frac{r_{+}}{2}+\frac{Q^{2}}{2r_{+}}-\frac{\lambda Q^{4}}{40r_{+}^{5}}+\frac{4\pi P r_{+}^{3}}{3},\\
T&=2Pr_{+}+\frac{1}{4\pi r_{+}}-\frac{Q^{2}}{4\pi r_{+}^{3}}+\frac{\lambda Q^{4}}{16\pi r_{+}^{7}}.
\end{aligned}
\label{eq:app-thermodynamic-quantities}
\end{equation}
Each equilibrium curve is evaluated at fixed $(P,Q,\lambda)$, while different values of $\lambda$ are considered to examine the changes in the phase structure.

Using $S=\pi r_+^2$, the Gibbs free energy is
\begin{equation}
\begin{aligned}
G=M-TS=\frac{r_{+}}{4}+\frac{3Q^{2}}{4r_{+}}
-\frac{2\pi P r_{+}^{3}}{3}
-\frac{7\lambda Q^{4}}{80r_{+}^{5}}.
\end{aligned}
\label{eq:app-gibbs}
\end{equation}
At a fixed temperature, the black hole branch with the lowest Gibbs free energy is thermodynamically preferred.

The nonlinear electromagnetic contribution to $T$ scales as $r_+^{-7}$ and becomes increasingly significant at small horizon radii. For positive $\lambda$, the temperature function behaves as $T\sim\lambda Q^4/(16\pi r_+^7)$ as $r_+\to0^+$. Depending on the parameters, the resulting temperature curve may exhibit additional branches and turning points.

\begin{figure}[htp!]
\centering
\includegraphics[width=\linewidth]{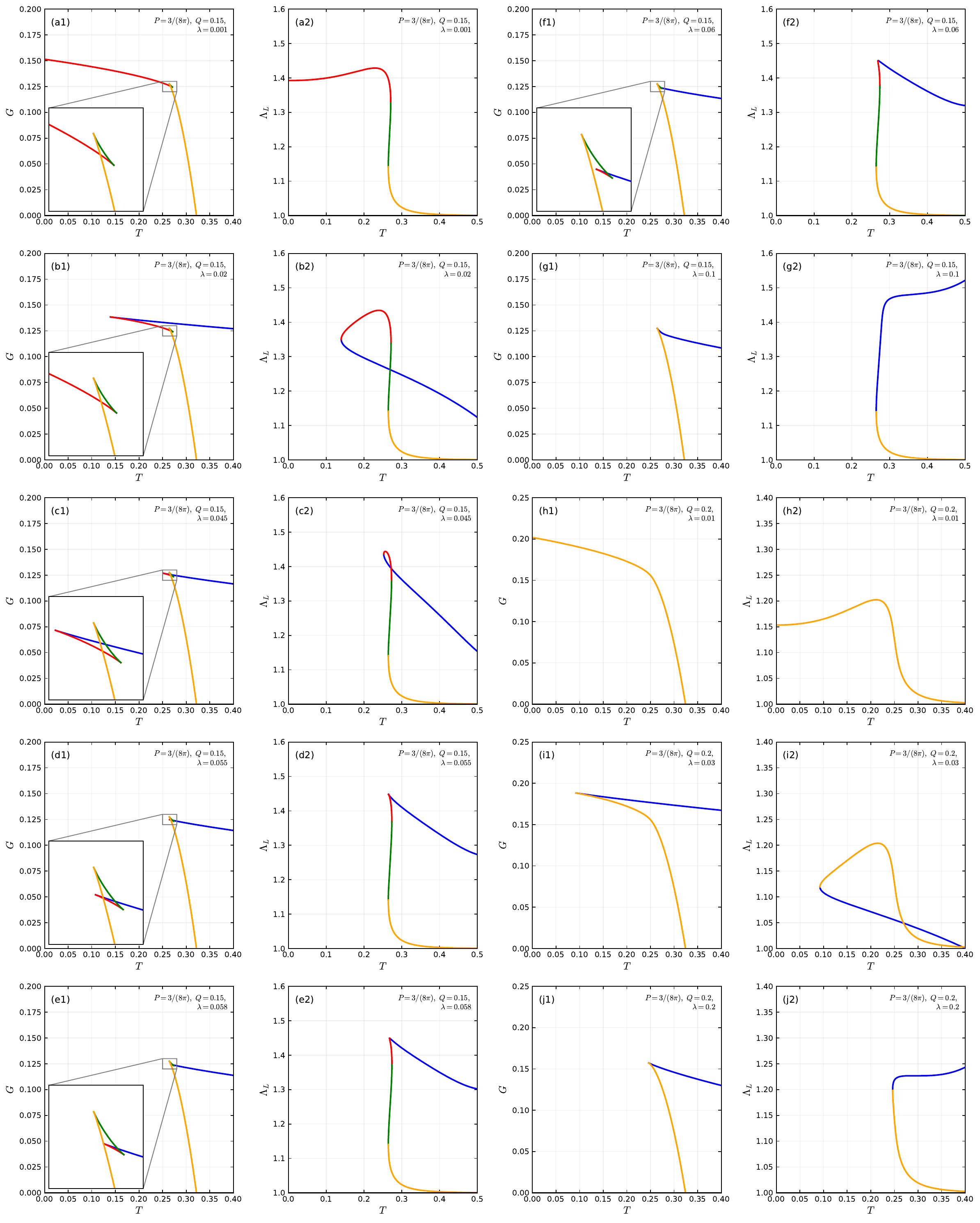}
\caption{Gibbs free energy $G$ and photon-sphere Lyapunov exponent $\Lambda_L$ as functions of the Hawking temperature $T$ for representative EEH-AdS black hole configurations. The curves are parametrized by $r_+$ at fixed $P$, $Q$, and $\lambda$, with $l=1$ and $P=3/(8\pi)$.}
\label{fig:EEHAdSChargeM}
\end{figure}

Figure~\ref{fig:EEHAdSChargeM} shows the thermodynamic and Lyapunov-exponent curves for representative values of $Q$ and $\lambda$. Depending on the parameters, the temperature admits a zero-temperature limit or a finite minimum. The Gibbs free-energy curves exhibit different multibranch structures, including first-order small/large-black-hole transitions and, in certain parameter regions, an additional zeroth-order transition.

The additional small-$r_+$ branch appears in the high-temperature region. Its Gibbs free energy may lie above that of competing branches or become the lowest over a finite temperature interval, depending on the parameters. The large-radius branch exhibits comparatively smaller changes with $\lambda$. Corresponding multivalued structures are also observed in $\Lambda_L(T)$.

To calculate the photon-sphere Lyapunov exponent, we consider null geodesics in the equatorial plane $\theta=\pi/2$. The geodesic Lagrangian is
\begin{equation}
\mathcal{L}_{\mathrm{geo}}
=\frac{1}{2}g_{\mu\nu}\dot{x}^{\mu}\dot{x}^{\nu}
=\frac{1}{2}\left[
-f(r)\dot{t}^{\,2}
+\frac{\dot{r}^{\,2}}{f(r)}
+r^{2}\dot{\phi}^{\,2}
\right],
\label{eq:app-geodesic-lagrangian}
\end{equation}
where a dot denotes differentiation with respect to an affine parameter. The conserved energy and angular momentum are
\begin{equation}
E=f(r)\dot{t},
\qquad
L=r^{2}\dot{\phi}.
\label{eq:app-geodesic-conserved}
\end{equation}
Using the null condition $\mathcal{L}_{\mathrm{geo}}=0$, the radial equation becomes
\begin{equation}
\dot{r}^{\,2}+\mathcal{V}_{\mathrm{eff}}(r)=0,
\qquad
\mathcal{V}_{\mathrm{eff}}(r)
=-E^{2}+\frac{L^{2}}{r^{2}}f(r).
\label{eq:app-null-effective-potential}
\end{equation}

For a circular null orbit, the effective potential satisfies
\begin{equation}
\mathcal{V}_{\mathrm{eff}}(r_{\mathrm{ph}})=0,
\qquad
\mathcal{V}_{\mathrm{eff}}'(r_{\mathrm{ph}})=0.
\label{eq:app-photon-orbit-conditions}
\end{equation}
These conditions yield
\begin{equation}
r_{\mathrm{ph}}f'(r_{\mathrm{ph}})
-2f(r_{\mathrm{ph}})=0.
\label{eq:app-photon-sphere}
\end{equation}
We select an unstable circular null orbit outside the event horizon, satisfying $r_{\mathrm{ph}}>r_+$ and $\mathcal{V}_{\mathrm{eff}}''(r_{\mathrm{ph}})<0$.

For a small radial displacement $\delta r$ from the circular orbit, linearizing the radial equation gives
\begin{equation}
\frac{\mathrm{d}^{2}\delta r}{\mathrm{d}\tau^{2}}
=-\frac{1}{2}
\mathcal{V}_{\mathrm{eff}}''(r_{\mathrm{ph}})\delta r,
\end{equation}
where $\tau$ is the affine parameter. The exponential growth rate measured with respect to the coordinate time $t$ defines the Lyapunov exponent,
\begin{equation}
\Lambda_L
=\left.
\sqrt{-\frac{\mathcal{V}_{\mathrm{eff}}''(r)}
{2\dot{t}^{\,2}}}
\right|_{r=r_{\mathrm{ph}}}.
\label{eq:app-lyapunov-general}
\end{equation}
Substituting the conserved quantities and using the circular-orbit conditions, we obtain
\begin{equation}
\Lambda_L
=\sqrt{
\frac{f(r_{\mathrm{ph}})
\left[
2f(r_{\mathrm{ph}})
-r_{\mathrm{ph}}^{2}f''(r_{\mathrm{ph}})
\right]}
{2r_{\mathrm{ph}}^{2}}
}.
\label{eq:app-lyapunov}
\end{equation}

For each black hole configuration, $r_{\mathrm{ph}}$ is determined from Eq.~\eqref{eq:app-photon-sphere}, and the corresponding $\Lambda_L$ is evaluated using Eq.~\eqref{eq:app-lyapunov}. As shown in Fig.~\ref{fig:EEHAdSChargeM}, different branches generally have different Lyapunov exponents, and the multibranch structure of $T(r_+)$ is also reflected in $\Lambda_L(T)$.

\begin{figure}[htp!]
\centering
\includegraphics[width=\linewidth]{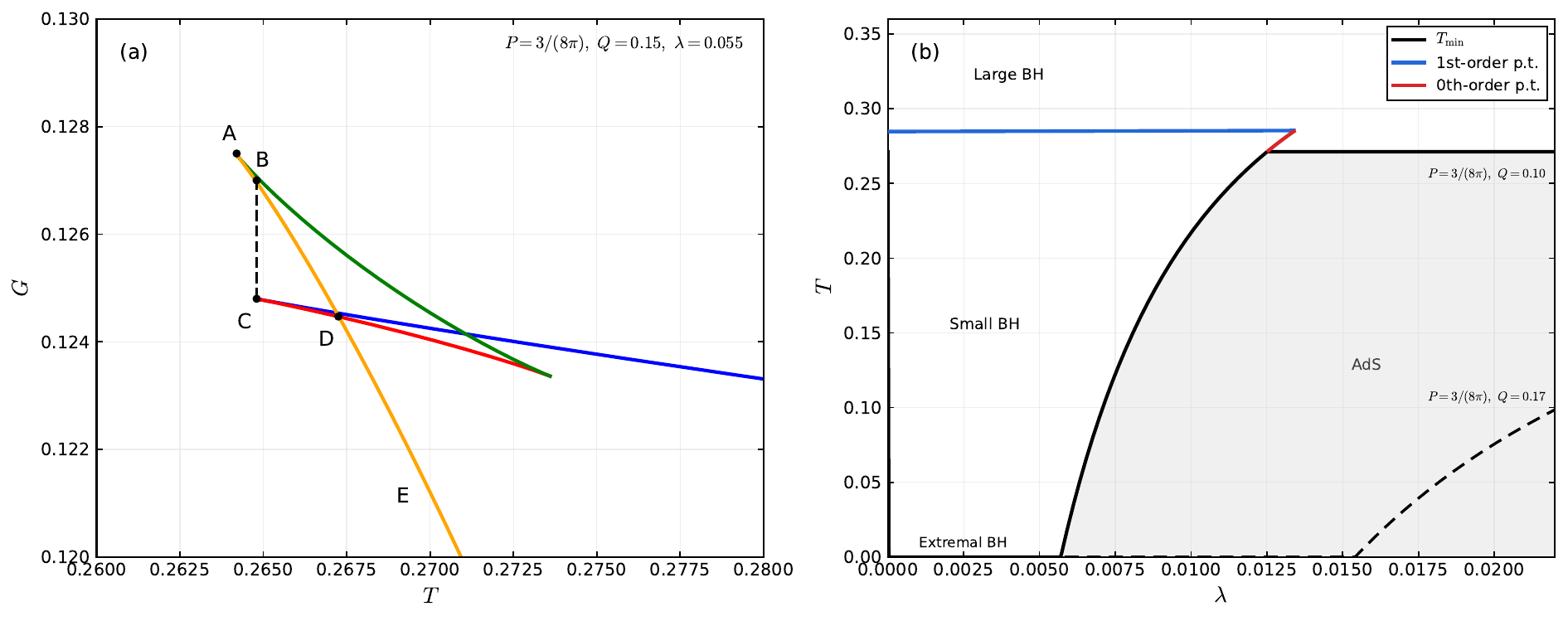}
\caption{Thermodynamic phase structure of the EEH-AdS black hole. (a) Gibbs free energy $G$ as a function of $T$ for $P=3/(8\pi)$, $Q=0.15$, and $\lambda=0.055$. (b) $T$-$\lambda$ phase diagram for $P=3/(8\pi)$, with $Q=0.10$ and $Q=0.17$.}
\label{fig:eeh-fixed-pq-phase-structure}
\end{figure}

A first-order phase transition occurs when two competing branches have the same temperature and Gibbs free energy,
\begin{equation}
G(r_{+,1})=G(r_{+,2}),
\qquad
T(r_{+,1})=T(r_{+,2}).
\label{eq:app-first-order-condition}
\end{equation}
The Gibbs free energy remains continuous across the transition, while the horizon radius and entropy change discontinuously.

A zeroth-order transition is characterized by a discontinuous change in the lowest Gibbs free energy. The corresponding free-energy gap is
\begin{equation}
\Delta G_0
=G(r_{+,2})-G(r_{+,1})\neq0.
\label{eq:app-zeroth-order-gap}
\end{equation}
In the examples considered here, such transitions occur when the thermodynamically preferred branch terminates and the preferred configuration changes to another branch.

Figure~\ref{fig:eeh-fixed-pq-phase-structure}(a) shows successive zeroth- and first-order transitions. As the temperature increases, the preferred branch evolves continuously from A to B. At the zeroth-order transition, the preferred configuration changes from B to C, accompanied by a finite jump in the Gibbs free energy. The branch then evolves from C to D, where a first-order transition occurs between two configurations with equal temperature and Gibbs free energy. The preferred configuration subsequently follows the branch toward E. The transition sequence is
\begin{equation}
\mathrm{A}\longrightarrow\mathrm{B}
\overset{\mathrm{0th}}{\longrightarrow}
\mathrm{C}\longrightarrow\mathrm{D}
\overset{\mathrm{1st}}{\longrightarrow}
\mathrm{E}.
\label{eq:app-phase-transition-sequence}
\end{equation}

Figure~\ref{fig:eeh-fixed-pq-phase-structure}(b) shows the phase structure in the $T$-$\lambda$ plane. For $P=3/(8\pi)$ and $Q=0.10$, the black solid curve represents the minimum-temperature boundary $T_{\min}(\lambda)$. The blue and red curves denote the first- and zeroth-order transition temperatures, respectively. Where both transition lines are present, increasing the temperature at fixed $\lambda$ first crosses the zeroth-order transition and then the first-order transition.

For $Q=0.17$, the black dashed curve gives the corresponding minimum-temperature boundary. No first- or zeroth-order transition lines appear within the displayed parameter range. Thus, the two charge configurations exhibit different phase structures, although both show a minimum-temperature boundary.

\end{document}